\documentclass[lettersize,journal]{IEEEtran}
\usepackage{amsmath,amsfonts}
\usepackage{algorithm}
\usepackage{algpseudocode}

\usepackage{array}

\usepackage{color}
\usepackage{booktabs}
\usepackage{amsmath}
\usepackage{arydshln}
\usepackage{subfigure}
\usepackage{float}
\usepackage{tcolorbox}
\usepackage{threeparttable}
\usepackage{colortbl}
\usepackage{multirow}
\usepackage{stmaryrd} 

\usepackage{amssymb}
\usepackage{booktabs}
\usepackage[linesnumbered,ruled,vlined,algo2e]{algorithm2e}

\usepackage{subfigure}
\usepackage{svg}
\usepackage{xstring}

\usepackage{algorithm}

\usepackage{url}
\usepackage{algpseudocode}
\SetKwInOut{Input}{}   
\SetKwInOut{Output}{Output}
\SetAlgoNlRelativeSize{-1} 
\SetAlgoNlRelativeSize{-1} 
\SetNlSty{}{}{}  

\usepackage{threeparttable}     

\usepackage{framed}
\usepackage{verbatim}
\usepackage{listings}
\usepackage{xcolor}

\definecolor{darkgreen}{rgb}{0.0, 0.5, 0.0}
\definecolor{darkred}{rgb}{0.5, 0.0, 0.0}
\lstdefinelanguage{Verilog}{
    morekeywords=[1]{module, begin,end,endmodule, always, posedge, assert}, 
    morekeywords=[2]{input, output, wire, reg,}, 
    keywordstyle=[1]\color{darkred}\bfseries,
    keywordstyle=[2]\color{darkgreen}\bfseries,
    sensitive=false,
    morecomment=[l]{//},
    morecomment=[s]{/*}{*/},
    morestring=[b]",
}

\lstdefinestyle{verilog-style}{
    language=Verilog,
    basicstyle=\ttfamily\footnotesize,
    commentstyle=\color{blue}\ttfamily\small,
    stringstyle=\color{darkred}\ttfamily\small,
    breaklines=true,
    showstringspaces=false,
    tabsize=4
}
\usepackage{multirow}

\usepackage{subfigure}
\usepackage{float}
\usepackage{makecell}
\usepackage{tabularx} 
\usepackage{hyperref}
\usepackage{graphicx}
\usepackage{xspace}
\usepackage{tcolorbox}

\usepackage{graphicx}
\usepackage{subcaption}
\usepackage{graphicx}
\usepackage{subcaption}
\usepackage{tikz}

\usepackage{ragged2e}
\usepackage{textcomp}
\usepackage{stfloats}
\usepackage{seqsplit}
\usepackage{threeparttable}
\providecommand{\Description}[1]{}

\usepackage{xcolor}
\usepackage{multirow} 
\usepackage{booktabs}
\usepackage{verbatim}
\usepackage{graphicx} 
\usepackage{subfigure}
\usepackage{listings}
\def\BibTeX{{\rm B\kern-.05em{\sc i\kern-.025em b}\kern-.08em
    T\kern-.1667em\lower.7ex\hbox{E}\kern-.125emX}}
\usepackage{balance}
\begin{document}
\title{A Novel Mutation Based Method for Detecting FPGA Logic Synthesis Tool Bugs}
\author{ Yi Zhang, He Jiang, Xiaochen Li, Shikai Guo, Peiyu Zou and Zun Wang
\thanks{He Jiang is with the School of Computer Science , Beijing Institute of Technology, Beijing 100081, China, and also with the School of Software, Dalian University of Technology, Dalian 116024, China, (corresponding e-mail: jianghe@dlut.edu.cn).

Yi Zhang is with the School of Computer Science, Beijing Institute of Technology, Beijing 100081, China (e-mail: zhangyi135336@163.com).

 Xiaochen Li, Peiyu Zou and Zun Wang are with the School of Software, Dalian University of Technology, Dalian 116024, China, and also with the Key Laboratory for Ubiquitous Network and Service Software of Liaoning Province, Dalian 116024, China (e-mail: xiaochen.li@dlut.edu.cn; zoupeiyu@mail.dlut.edu.cn; wangzun\_ssdut2020@mail.dlut.edu.cn).

Shikai Guo is with the School of Information Science and Technology, Dalian 
Maritime University, Dalian 116026, China (e-mail: shikai.guo@dlmu.edu.cn).

}}


\maketitle

\begin{abstract}
FPGA (Field-Programmable Gate Array) logic synthesis tools are the key components in the EDA (Electronic Design Automation) toolchain, which convert hardware designs written in description languages (e.g., Verilog) into gate-level representations for FPGAs. However, defects in these tools may lead to unexpected behaviors and pose security risks. Therefore, it is crucial to harden these tools through testing. Although several methods have been proposed to automatically test FPGA logic synthesis tools, the challenge still need to be addressed, namely the insufficient semantic logical complexity of test program. In this paper, we propose VERMEI, a new  method for testing FPGA logic synthesis tools. VERMEI consists of three modules: preprocessing, equivalent mutation, and bug identification. The preprocessing module identifies zombie logic (inactive code with no impact on the circuit’s output) in seed programs through simulation and coverage analysis. 
The equivalent mutation module generates the equivalent variants  of seed programs by pruning or inserting logic fragments in zombie areas. It uses Bayesian sampling to extract logic fragments from historical Verilog designs for insertion, making the generated equivalent variants have complex control flows and structures, addressing the challenge. The bug identification module, based on differential testing, compares the synthesized outputs of seed and variant programs to identify bugs. Experiments over Yosys, Vivado, and Quartus demonstrate that VERMEI outperforms the state-of-the-art methods in testing FPGA logic synthesis tools. Within five months, VERMEI reported 15 bugs to  vendors, 9 of which were confirmed as new bugs.
\end{abstract}


\section{Introduction}
FPGA synthesis tools are used to automatically convert high-level design descriptions written in hardware description languages (such as Verilog ) into gate-level netlists that can be implemented on FPGAs. They serve as a critical link between theoretical hardware design and the actual circuit implementation in EDA \cite{koblah2023survey,sharma2016high,tu2024logic,ling2015accelerating,cong2014minimizing}. 
These tools play a vital role in enabling digital computation and building reliable hardware devices. For instance, the design and implementation of many modern digital systems, including microprocessors, accelerators, and embedded devices, depend on FPGA logic synthesis tools to ensure efficient and accurate hardware realization \cite{Intel,ling2015accelerating}. 
 Like other application software, FPGA logic synthesis tools are prone to  bugs \cite{tu2022detecting,tang2021detecting,guo2022detecting,you2023regression,jiang2024testing}. These bugs can affect the accuracy of hardware design and may manifest indirectly as FPGA failures, which can lead engineers to mistakenly attribute these hardware failures to bugs within the hardware models themselves rather than to the FPGA logic synthesis tools.
 
To this end, several methods have been proposed to generate random Verilog test programs for testing FPGA logic synthesis tools. The first method, VlogHammer, was introduced by Clifford Wolf~\cite{VlogHammer}, which defines and samples a subset of Verilog syntax to generate test programs. However, it cannot generate HDL code with multiple modules or behavioral-level constructs like “always” blocks. Verismith~\cite{Verismith} was later proposed as an extension of VlogHammer,uses an Abstract Syntax Tree (AST)-based approach to generate pseudo-random and valid Verilog test programs. However, Verismith is limited in its ability to generate semantically complex test programs, and the deterministic generation patterns often fail to explore deeper corner cases in synthesis tools. EvoHDL~\cite{ledohdl} uses the fixed CPS models to transform into various types of HDL programs (i.e., Verilog, VHDL, and SystemVerilog) for testing FPGA logic synthesis tools. However, although the CPS models are converted into different HDL types to create diverse test inputs, the limited module set and the fixed control and data flow patterns constrain the diversity and complexity of the generated programs’ structural and semantic features. Therefore, despite the fact that existing methods have helped developers detect bugs in FPGA logic synthesis tools to some extent, the test programs they generate still have significant room for improvement in bug detection capability.


The complexity and variability of real-world FPGA synthesis workflows present major challenges for testing FPGA logic synthesis tools.

\textbf{Challenge: Insufficient Semantic Logical Complexity.} Current state-of-the-art Verilog test program generators typically cover only a limited subset of combinatorial patterns within syntax structures. Specifically, Verismith generates test cases by randomly sampling the Verilog syntax specification, while EvoHDL generates structurally valid HDL programs by systematically combining a fixed set of CPS (Continuation-Passing Style) models. Due to the inherent limitations of these generation strategies and configurations, the resulting test programs often have simple structures, containing only basic combinational patterns. This makes it difficult to simulate the diversity and complexity of industrial-scale designs. As a result, such test programs are not effective in uncovering deeper bugs in FPGA logic synthesis tools. Therefore, a key challenge is how to generate Verilog test programs with higher semantic logic complexity to more closely align with the needs of real industrial designs, and thus to test FPGA logic synthesis tools more thoroughly.	


In this paper, we propose VERMEI (\textbf{VER}ilog \textbf{M}utation based on \textbf{E}quivalent \textbf{I}nput), an effective technique to generate Verilog test programs with diverse input patterns and higher complexity for detecting bugs in FPGA logic synthesis tools. VERMEI consists of three modules: preprocessing, equivalent mutation, and bug identification. Specifically, the preprocessing module first constructs a robust simulation environment (i.e., a testbench) based on the input variables (including clock signals) of the seed Verilog test program. It then performs simulation and coverage analysis to identify zombie logic (Section~\ref{sec:Zombie Logic}) (inactive code with no impact on the circuit’s output) in the Verilog program.  Next, the equivalent mutation module generates equivalent variants of the seed test program by iteratively trimming or inserting new logic fragments in the zombie logic areas. During the insertion process, the equivalent mutation module uses a Bayesian mechanism to sample candidate logic fragments with representative and differentiated characteristics from historical Verilog circuit designs. This process incorporates industry-standard scenarios and various Verilog structural features, such as complex control flow structures (e.g., while, case, and forever loops), nested conditional statements, and intricate computation structures. In this way, the equivalent variants with higher semantic and logical complexity are generated, allowing us to overcome the \textbf{Challenge}. Finally, the bug identification module uses differential testing to validate the generated Verilog variants. It compares the outputs of the seed program and its variants after synthesis  under the same stimuli. If a variant's output differs from the seed program, it is considered to trigger a bug.

To assess the effectiveness of  VERMEI, we conducted the evaluation over the FPGA logic synthesis tools. 
First, we compare VERMEI against the state-of-the-art methods (i.e., Verismith and EvoHDL). The results demonstrates that VERMEI achieved a 100\% improvement in bug detection effectiveness compared to Verismith and a 60\% improvement compared to EvoHDL. 
Within five months, we reported a total of 15 bugs in the latest versions of various FPGA logic synthesis tools, of which 9 bugs have been confirmed as new bugs. 
Second, we examine the influence of mutation operations. 
Using the variants of VERMEI\textsubscript{\textit{insert}} (i.e., VERMEI with only inserting operators) and VERMEI\textsubscript{\textit{prune}} (i.e., VERMEI with only pruning operators), VERMEI outperforms its variants by detecting 14.69\% $\sim$ 34.29\% more bugs on average. Finally, we conduct a comparative experiment that compares VERMEI with the baselines (Verismith and LgooHDL) in terms of test coverage. VERMEI achieves higher coverage than Verismith by 9.6\% in line, 9.9\% in condition, and 7.52\% in branch metrics; and outperforms EvoHDL by 8.14\%, 8.74\%, and 7.1\%, respectively.

In summary, this paper makes the following contributions:

\begin{itemize}
 \item We propose VERMEI, a novel testing method for detecting bugs in FPGA logic synthesis tool. VERMEI utilizes the equivalent mutation module and the bug identification module to address the insufficient semantic logic complexity challenge in FPGA logic synthesis tool testing.
  \item We conducted extensive experiments on Yosys, Vivado, and Quartus Prime to evaluate the bug detection capability of VERMEI. VERMEI has identified 15 valid bugs, 9 of which have been confirmed by the developers as new bugs.
 \item We have open-sourced the code of VERMEI  on Figshare\footnote{https://figshare.com/s/17b24362c3258f05e942}.
 
\end{itemize}

\section{Background}
\subsection{Verilog and the process of FPGA Logic Synthesis}

\begin{figure}[!t]
	\centering
\includegraphics[width=0.99\linewidth]{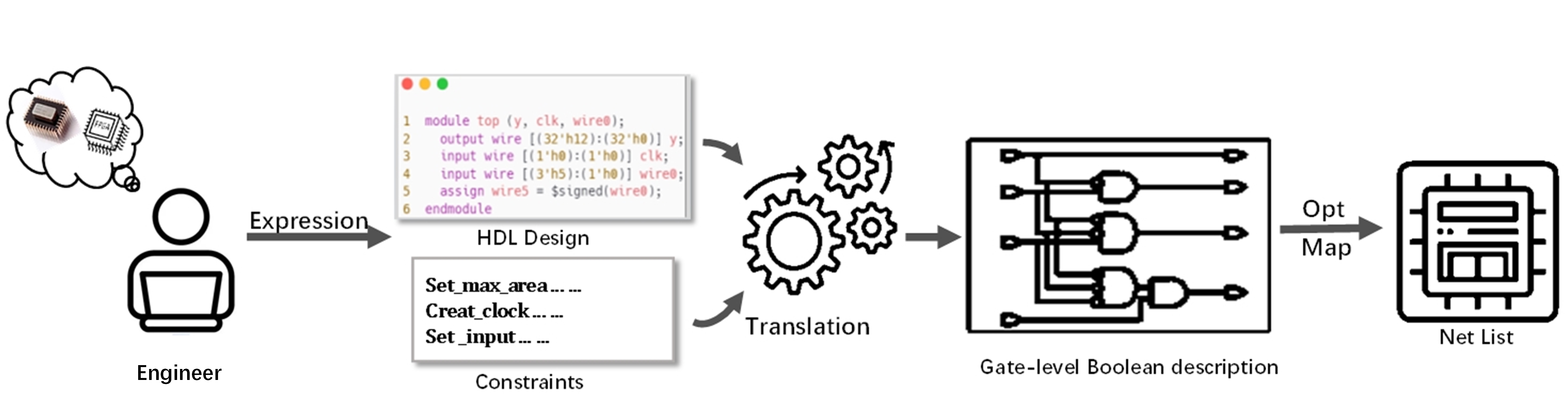}
	\caption{The process of FPGA logic synthesis
}
 \Description{}
  \label{fig:synthesis}
\end{figure}

\textit{Verilog} is a Hardware Description Language (HDL) commonly used for designing, simulating, and verifying digital circuits. It includes a subset of constructs that can be directly synthesized into hardware, enabling designs to be implemented on devices such as FPGAs. Unlike procedural programming languages (e.g., C or Python), Verilog supports parallel and clock-driven execution, which allows designers to describe both the structural and behavioral aspects of hardware systems.
This contrasts with the linear execution model of procedural languages, where operations are typically performed sequentially. Verilog, in contrast, is timing-sensitive—logic behavior depends not only on the current signal states but also on their transitions across clock cycles. Even if certain logic is inactive in one cycle, it may still be triggered in subsequent ones.

\textit{FPGA logic synthesis} is the process of transforming hardware designs described in hardware description languages (e.g., Verilog) into low-level components, such as gate-level netlists, that can be efficiently mapped and deployed onto an FPGA \cite{hachtel2007logic, mondal2020ising}. 
FPGA logic synthesis tools such as Yosys \cite{yosys}, Vivado \cite{Xilinx} or Quartus \cite{Quartus} 
enable designers to focus on the algorithmic and architectural aspects of their designs, while automating the conversion of abstract specifications into practical logic circuits that can be implemented on FPGAs. By designs optimizing for speed, area, and power, these tools enhance design productivity and streamline FPGA development.
\figureautorefname~\ref{fig:synthesis} illustrates the FPGA logic synthesis process. Designers first use hardware description languages (HDLs), such as Verilog, to specify the circuit's functionality and structure, including logic operations, timing, and data flow \cite{rai2021logic}. Guided by user-defined constraints (e.g., timing, area, and power), the synthesis tool analyzes the HDL code and transforms it into an optimized gate-level representation mapped to specific hardware components. After synthesis, the design undergoes verification—such as timing analysis and functional simulation—to ensure correctness and performance compliance. Finally, during placement and routing, logic elements are assigned physical locations and interconnections, producing a bitstream file that configures the FPGA to execute the intended functionality.



Therefore, the synthesis process is critical as it bridges the gap between the theoretical design and its actual physical implementation on the FPGA hardware. Any defect in the FPGA logic synthesis tools can lead to unexpected behavior in the final FPGA device, potentially resulting in significant performance issues or safety risks.

\subsection{Zombie Logic}
\label{sec:Zombie Logic}

\begin{figure}[!t]
	\centering
\includegraphics[width=0.92\linewidth]{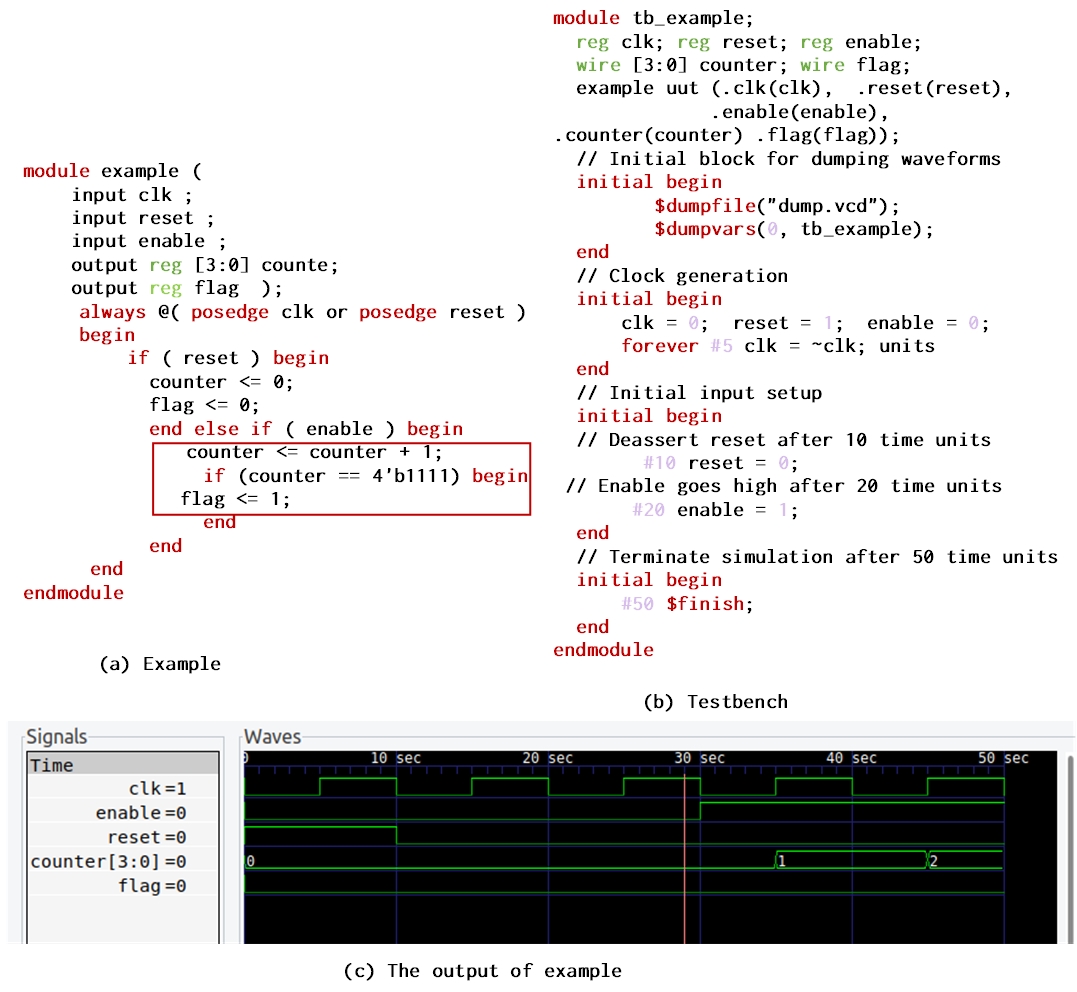}
	\caption{Example module demonstrating counter behavior with reset and enable signals
}
 \Description{}
  \label{fig:example_module}
\end{figure}

Handling of conditional execution within logic blocks has been a research challenge, especially in data flow models \cite{bainomugisha2013survey}.  
While conditional execution is easily understood in procedural programming, in HDL (e.g., Verilog), its execution is strictly constrained by clock-edge triggering and synchronous signal transitions of input signals. This timing dependency makes the conditional execution of logic blocks in HDL different from the linear control flow of procedural languages, where the state may switch dynamically at any clock cycle rather than being determined by a static code sequence. This fundamental difference makes it impossible to map “dead code” (i.e., code in a program that is never executed or has no effect on the output) from traditional software engineering directly to the hardware design domain.

To more accurately characterize logic that is structurally present but behaviorally inactive, we introduce the notion of Zombie Logic, which is a block of logic that is temporarily inactive (i.e., not executing or having no effect on the current output) under certain clock cycles or input conditions, but may become reactivated due to future signal transitions or timing state changes. Its state may change to active due to synchronization events (e.g., signal jumps or timing conditions are met), reflecting an intermediate semantic state between software “dead code” and “don't care” in Boolean logic.
 
Meanwhile, based on whether a logic block can be activated under any timing condition, Zombie Logic can be further categorized into Static Zombie Logic and Dynamic Zombie Logic. Static Zombie Logic corresponds to “dead code” in Boolean circuits (e.g., an \textit{if (0)} branching), i.e., the logic is not reachable under all input conditions and clock cycles. This type of logic can be recognized and safely removed by synthesis tools. Dynamic Zombie Logic is logic that is unreachable in the current input state or clock cycle, but may be activated under certain history paths or future signal jumps. For example, in \figureautorefname ~\ref{fig:example_module}(a), the logic \textit{counter <= counter + 1} inside the \textit{else if (enable)}  branch is skipped when \textit{enable = 0}, but can become active again when \textit{enable = 1} in subsequent cycles. The retention mechanism is similar to “don't care” in Boolean optimization, i.e., it currently has no effect on the output, but potentially has an effect on the timing behavior and cannot be removed at will.

\section{Framework  of VERMEI} 
\label{sec:apporach}
\begin{figure*}[!t]
	\centering
\includegraphics[width=0.98\linewidth]{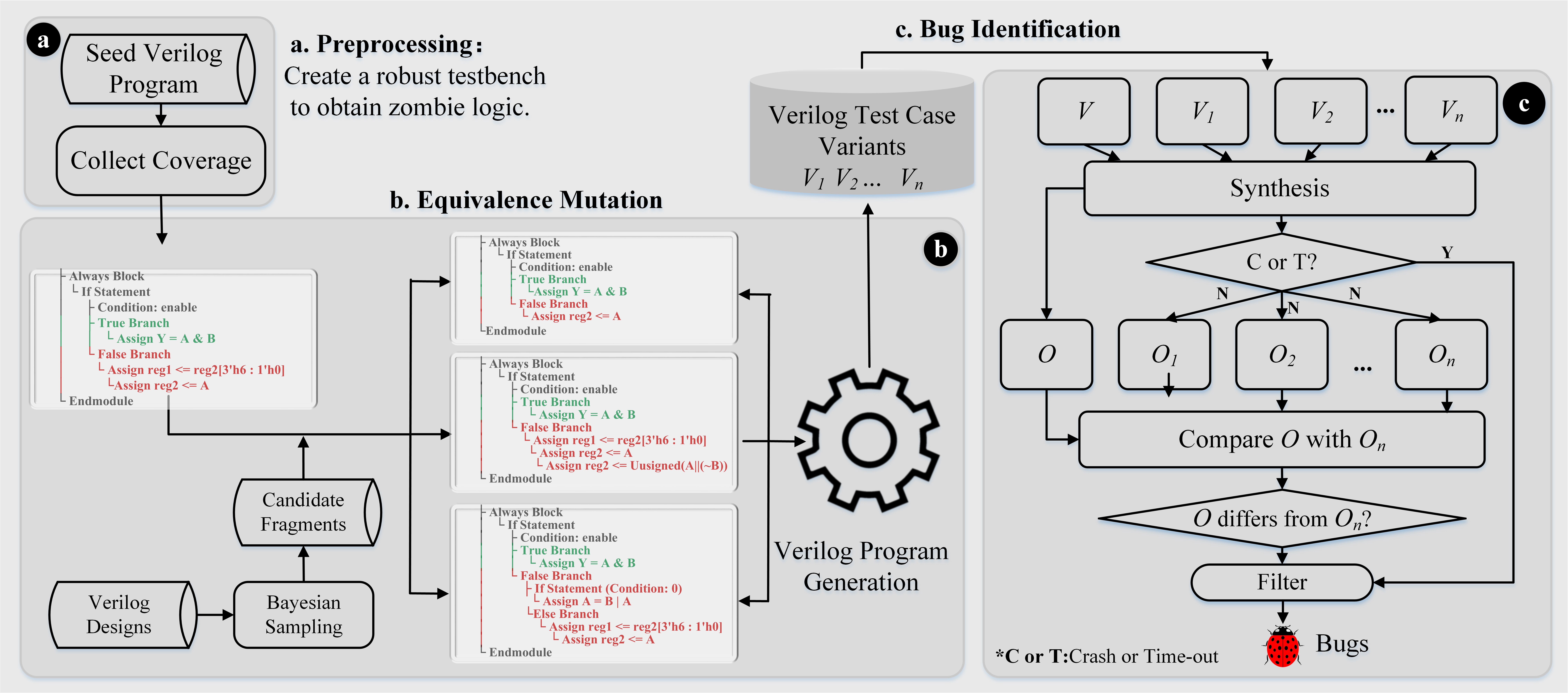}
	\caption{The framework of VERMEI}
  \Description{}
 \label{fig:The framework of VERMEI}
\end{figure*}

In this section, we describe the design of our proposed
VERMEI framework. 
\figurename ~\ref{fig:The framework of VERMEI} and Algorithm 1 show the overall workflow of VERMEI, which includes three parts, namely preprocessing, equivanlence mutation, and bug identification. 
The first part aims to annotate zombie logic. 
The second part involves the generation of equivalent variants from the seed Verilog test program. The third part applies the differential testing strategies  to identify potential bugs. 

As aforementioned in Section~\ref{sec:Zombie Logic}, one of the most challenging aspects of equivalence-based input mutation for testing FPGA logic synthesis tools is identifying instances of ‘zombie’ logic in Verilog.
We acknowledge that the Verilog language exhibits characteristics similar to procedural languages within each \textit{always} block; however, Verilog executes concurrently and incorporates a well-defined timing concept. 
Even if a particular logic does not affect the output under the current inputs and state, it may contribute under different timing conditions or states. 
Additionally, a currently inactive logic branch may still be part of the Verilog design due to its association with clock edges or other synchronous events. 
In this study, our objective is to modify zombie logic, which does not impact hardware behavior across any clock cycle or input-state combination, to generate equivalent variants. This method aims to expose deeper bugs in FPGA logic synthesis tools. 
To achieve this, we utilize coverage tools such as VCS \cite{vcs} to identify zombie logic and construct a robust testing environment (testbench) based on input-output characteristics from seed Verilog test programs (algorithm 1, lines 2-5). 
For the same time, VERMEI generates equivalent variants by pruning or inserting logic fragments in zombie areas.
The inserted fragments are generated through Bayesian sampling, based on historical data and existing patterns, ensuring compatibility with the seed Verilog test program's context (\sectionautorefname~\ref{equivalence Mutation}).  
With the assistance of effective Verilog variants, we conduct  differential testing strategies to identify bugs (lines 8-11).  If the output of a Veriolg variant is different from the seed Veriolg program, it is deemed to trigger a FPGA logic synthesis tool bug.

\begin{algorithm}[!t]
\small
\caption{VERMEI}
\raggedright
\KwIn{Logic Synthesizer $S$,  Verilog test program $V$, Testbench $T$}
\KwOut{Reported Bugs ${B}$}
\SetKwFunction{FMain}{TEST}
\SetKwProg{Fn}{Procedure}{:}{}
\Fn{\textbf{\texttt{TEST}} ( $S$,  $V$,  $T$)}
{
   \tcp{Step 1: Extract coverage and V\_Input }
    \textit{V\_Input} $\gets$ \textit{Verilog test program } $V$\;
    
    \textit{Testbench} $T$ $\gets$ \textit{V\_Input} $\gets$ \textit{random}(0...N)\;
    
    $V_{exe} \gets S.\textit{Synthesize}(V, T)$ \;
    
    $C \gets \{\textit{Coverage}(V_{exe}.\textit{Execute}(i)) \mid i \in T\}$ \;
    
    \tcp{Step 2: Generate equivalent variants}
    \For{$\textit{iter} \gets 1$ \KwTo \textit{MAX-ITER}}{
        $V' \gets \textit{Equivalence Mutation}(V, T)$ \;
       
        \tcp{Step 3: Bug identification}
        $V'_{exe} \gets S.\textit{Synthesize}(V', T)$ \;
        
       \If{ $V_{exe} \neq V'_{exe}$ }{
            ${B} \gets {B} \cup \{(V, V', T)\}$ 
        }
    }
\Return ${B}$    
}
\end{algorithm}

\subsection{Preprocessing}           
Given a seed Verilog test program, VERMEI  annotates zombie logic (\ref{sec:Zombie Logic}). within the test program. VERMEI effectively marks these zombie logic elements without altering the Verilog test program's behavior, which is crucial for constructing equivalence variants. Specifically, to identify zombie logic, VERMEI simulates the original Verilog test program using an automatically generated testbench. For each seed file, it extracts the names, widths, and signedness of input/output ports, then assigns random values to the inputs while maintaining a fixed clock signal toggled periodically. The testbench stimulates the design under varying delays (e.g., \#10, \#20, ...) and terminates after a fixed simulation period. This randomized input strategy enables broad path exploration. VERMEI then uses built-in coverage tools (e.g., VCS) to monitor signal transitions and execution frequency, allowing precise identification of inactive (zombie) logic within the seed Verilog program.

 

\begin{algorithm}[t!]
\small 
\SetInd{1em}{1em} 
\raggedright
\setcounter{AlgoLine}{0}
\caption{Generation of Equivalent Variant}
\label{alg:equivalence_mutation}
\SetKwInOut{Input}{Input}
\SetKwInOut{Output}{Output}
\Input{Verilog test program $V$, Coverage $C$, Verilog logic statement $s$}   
\Output{Variant $V'$}
\SetKwProg{Fn}{Function}{:}{end}
\SetKwProg{Proc}{Procedure}{:}{}
\Proc{\textbf{\texttt{GenVariant}}(Verilog test program $V$, Coverage $C$)}{
    \ForEach{$s \in V$.\textit{Logics()}}{
        \eIf{FlipCoin()}{
            $V' \gets$ \textbf{\texttt{PruneVisit}}($V'$, $s$, $C$)\;
        }{
            $V' \gets$ \textbf{\texttt{InsertVisit}}($V'$, $s$, $C$)\;
        }
    }
    \textbf{\texttt{VerifyFunctionality}}($V'$)\;
    
    \Return $V'$\;
}
\Fn{\textbf{\texttt{PruneVisit}}(Verilog test program $V$, Verilog logic statement $s$, Coverage $C$)}{
    $ast \gets transform(V)$ \\
    $node \gets traversal(ast, s)$ \\
    \If{$node \notin C$ \textbf{and} FlipCoin($node$)}{
        \If{\texttt{IsLeaf}($node$)}{
            $ast' \gets \textit{delete}(node, ast)$
        }
        \Else{
            $ast' \gets \textit{DeleteSubtree}(node, ast)$
        }
        $V' \gets \textit{transform}(ast')$ \\
        \Return{$V'$}
    }
    \tcp{Otherwise, traverse $s$'s children}
    \ForEach{$s' \in s$.ChildrenLogics()}{
        \textbf{\texttt{PruneVisit}}($V$, $s'$, $C$)\;
    }
    \Return $V'$\;
}
\Fn{\textbf{\texttt{InsertVisit}}(Verilog test program $V$, Verilog logic statement $s$, Coverage $C$)}{
    $ast \gets \textit{transform}(V)$ \\  
    $node \gets \textit{traversal}(ast, s)$ \\  
    \If{$node \notin C$ \textbf{and} FlipCoin($node$)}{
        $supplied \gets \textit{extract}(context, V\_input, V\_output)$\\
        $insert\_num \gets \textit{size}(sample\_Mut)$\\
        $sample \gets \textit{sample}(supplied)_{i \gets \textit{random}(1 \ldots insert\_num)}$ \\
        $ast' \gets \textit{insert}(node, ast, sample)$ \\
        $V' \gets \textit{transform}(ast')$ \\
        \Return $V'$\;
    }
    \tcp{Otherwise, traverse $s$'s children}
    \ForEach{$s' \in s$.ChildrenLogics()}{
        \textbf{\texttt{InsertVisit}}($V$, $s'$, $C$)\;
    }
    \Return $V'$\;
}
\end{algorithm}

\subsection{ Equivalence Mutation}
 Algorithm 2 describes the process by which VERMEI generates an equivalent variant. 
 The equivalence mutation including two key operations: pruning and insertion within the regions of zombie logic. We use the function \textit{FlipCoin} to decide
stochastically whether a zombie logic should be removed or added (line 3).
 To ensure that all variants generated by VERMEI are syntactically correct  Verilog test programs, these Verilog test programs are processed by parsing them into ASTs. For example, when VERMEI chooses to prune a logic, it eliminates all tokens within its AST subtree.

\textbf{Pruning Mutation.}  Algorithm 2 (lines 9–21) describes the process VERMEI uses to generate equivalence variants by applying pruning mutations to the abstract syntax tree (AST) of the seed Verilog test program \(V\). The algorithm performs a breadth-first traversal of the AST and applies a probabilistic decision to each node using the \textit{FlipCoin(node)} function (lines 12–18). Two separate pruning probabilities are defined: \(p_\text{parent}^d\) for non-leaf (parent) nodes  and \(p_{\text{leaf}}^d\) for leaf nodes. These parameters control the likelihood of pruning different types of logic constructs. If a leaf node is selected for pruning, it is directly removed. If a parent node is selected, the entire subtree rooted at that node (including all its descendants) is removed. This distinction reflects the fact that pruning a parent node has a broader structural effect on the program than pruning a leaf. This mutation process preserves syntactic correctness and is constrained to ensure semantic equivalence. 

\begin{figure}[!t]
	\centering
	\includegraphics[width=1\linewidth]{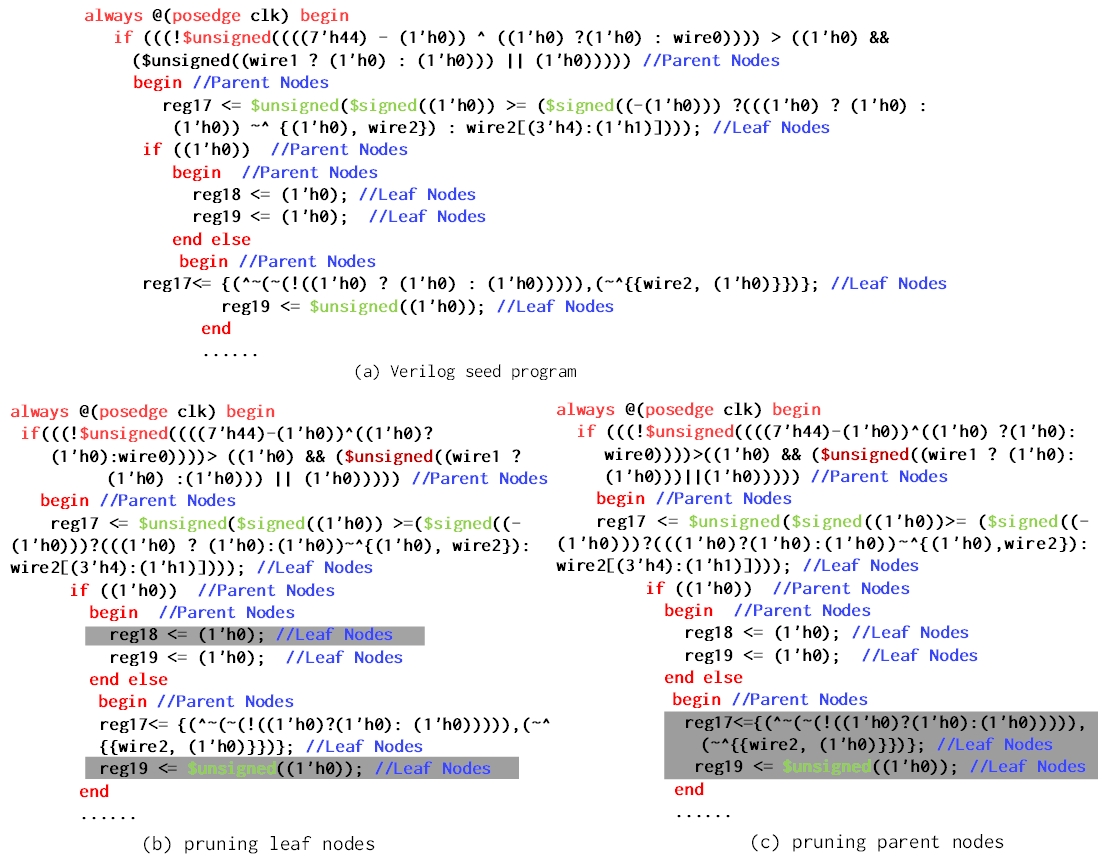}
	\caption{The example of pruning mutation}
\Description{}
 \label{fig:Examples_pruning}
\end{figure}

\figurename ~\ref{fig:Examples_pruning} presents a concrete example that illustrates how pruning mutations operate on different parts of the Verilog seed program. In \figurename ~\ref{fig:Examples_pruning}(a), the original Verilog program contains multiple nested control structures and assignments. We annotate the AST structure by highlighting the parent nodes (control structures such as \texttt{if-else} blocks) and the leaf nodes (individual assignment statements). When a \textbf{leaf node} is selected for pruning, as shown in Figure~4(b), only the specific assignment statements are deleted (e.g., \texttt{reg18 <= (1'h0);} and \texttt{reg19 <= (1'h0);} are removed), while the surrounding control structures (\texttt{if-else}) are preserved. This results in finer-grained modifications with relatively local impact on program behavior.
In contrast, when a \textbf{parent node} is pruned, as shown in \figurename~\ref{fig:Examples_pruning}(c), the entire subtree rooted at that control structure—including all its child statements—is deleted, as indicated by the gray shaded region, which leads to a more substantial structural modification.

However, Not all AST nodes in Verilog are suitable for pruning, as some deletions may introduce syntactic errors. For example, deleting a declared variable can cause errors in statements that depend on that variable. To avoid such issues, we prevent pruning of AST nodes associated with identifiers that are declared, defined, or referenced in the Verilog test program. To ensure syntactic correctness, we compile the pruned variant and check if an executable is generated. If compilation fails, the variant is discarded, and we attempt further pruning until a syntactically correct variant is produced.

\textbf{Inserting Mutation.} We also have two probabilities \(p_\text{parent}^d\) and \(p_{\text{leaf}}^d\) for inserting at parent or leaf logics (line 22-34). We can insert a new logic either before or after the existing logic, with an equal probability of selecting the insertion position. If the zombie logic does not directly belong to a \textit{begin-end} block, we promote it to a \textit{begin-end} block before inserting the new logic. This ensures that these logic elements share the same parent.

Additional, in the inserting mutation process, we maintain a database of candidate logic statements and select suitable ones for insertion. Utilizing a Bayesian sampling strategy, we generate logical fragments from existing Verilog designs. We mined over 24k Verilog designs\footnote{https://figshare.com/s/125133bde1e297b10684} by using automated scripts and FPGA tool APIs to extract and filter ".v” files from open-source repositories and FPGA logic synthesis tools. These files were then filtered to ensure they could be successfully compiled, which provided a diverse and high-quality dataset for analysis.
We utilize the Bayesian method to integrate prior knowledge with observed data, enhancing the accuracy of our reflections on real-world conditions. We also implement Bayesian sampling to dynamically update our model with new data, enabling the generated logical fragments to adapt more effectively to changing Verilog test program requirements. \cite{besag1995bayesian,kruschke2010bayesian,hoffman2014no}.

Given a Verilog design corpus \( \mathcal{V} \)  , we traverse the  AST of each design \( V \in  \mathcal{V}  \) to extract a set of syntax elements \( {Z} = \{z_1, z_2, \ldots, z_n\} \), where \( z_i \)  can represent either syntax operators (e.g., +, -, *) or control structures (e.g., if-else, case, for).  Let \( f(z_i) \) represent the frequency of syntax operator \( z_i \) in the corpus. We compute the complexity-weighted probability of occurrence for each syntax operator \( z_i \) as follows:
        \[
        P(z_i) = \frac{C(z_i) \cdot f(z_i)}{\sum_{j=1}^{n} C(z_j) \cdot f(z_j)}
        \]
where \( C(z_i) \) denotes the complexity weight assigned to \( z_i \), as detailed in Table~\ref{tab:complexity_weights_verilog}. And \(f(z_i)\) represents the frequency of syntax elements.

Let \( P(z_i \mid z_j) \) denote the conditional probability of syntax operator \( z_i \) given \( z_j \). Using Bayes' theorem, we can estimate the \( P(z_i \mid z_j) \) as flolows:
\[
P(z_i \mid z_j) = \frac{P(z_j \mid z_i) \cdot P(z_i)}{P(z_j)}
\]
where \( P(z_j \mid z_i) \) is the probability of \( z_j \) occurring given \( z_i \), and \( P(z_i) \) and \( P(z_j) \) are the individual occurrence probabilities of \( z_i \) and \( z_j \), respectively.

We apply this probabilistic model to guide the syntax-aware mutation process as follows. We first select a starting operator \( z_{\text{start}} \) based on its probability  \( P(z_{\text{start}}) \).  We then generate a sequence of syntax operators \( \{z_1, z_2, \ldots, z_k\} \) by sampling:

\[
z_{i+1} \sim P(z \mid z_i) \quad \text{for} \; i = 1, 2, \ldots, k-1
\]
 
 The acceptance criteria for the selected operators are defined by a threshold \( T \) based on the average or median occurrence frequency of less common operators. An operator \( z_i \) is selected if \( P(z_i \mid z_{i-1}) \geq T \), where \( z_{i-1} \) is the previously selected operator. 
 The sampling process terminates either when the sequence reaches a certain length \( L \) or meets specific end conditions (e.g., \textit{end}, \textit{endmodule}). To ensure syntactic correctness, we reuse variable and signal names from seed Verilog tests, keeping declarations and ports consistent. 
To  adapt to synthesis results, we incorporate a feedback-guided adjustment mechanism into our Bayesian sampling framework. After each mutation batch, we analyze the synthesis outcomes to identify which operators or structures contribute to activating new logic paths or improving coverage. We then adjust the complexity weights \( C(z_i) \) and the transition probabilities \( P(z_i \mid z_j) \), increasing the weights of elements correlated with successful mutations while decreasing the probabilities of less effective ones. This iterative feedback loop enables our sampler to prioritize mutations that are more likely to affect logic synthesis behavior.

 Finally, the generated code needs to pass basic syntax and logic checks to ensure its correctness. For example, port declarations in Verilog must be consistent with their use, all signals and variables must be declared before use, and so on. This sampling method allows us to generate and construct candidate logical fragments at the statement, block, and function levels effectively.

\begin{table}[!t]
\centering
\caption{Complexity Weights of Verilog Syntax Elements}
\label{tab:complexity_weights_verilog}
\small
\setlength{\tabcolsep}{2pt}
\begin{tabular}{p{3.2cm}|l|l|l}
\toprule
Type & Typical Operators & Number & Weight \\
\midrule
Unary Operator & + - ! \textasciitilde & 1 & 1 \\
Add, Subtract, Multiply, Divide, Modules & + - * / \% & 2 & 2 \\
Logical Operations & \&\& \textbar\textbar & 2 & 2 \\
Bitwise & \& \^ & 2 & 2 \\
Equivalence Operator & == != & 2 & 2 \\
Equivalence Operator & === !== & 2 & 3 \\
Relation & \texttt{<} \texttt{<=} \texttt{>} \texttt{>=} & 2 & 3 \\
Shift Operations & \texttt{<<} \texttt{>>} & 2 & 3 \\
Conditional Operators & ?: (Ternary) & 3 & 4 \\
Logical Control Structures & if-else, case & Multiple & 4 \\
Loop Structures & for, while & Multiple & 5 \\
\bottomrule
\end{tabular}
\end{table}

In VERMEI, pruning and insertion behaviors are controlled by two parameters, \(p_\text{parent}^d\) and \(p_{\text{leaf}}^d\) , which specify the probabilities to prune parent or leaf
statements in \textit{FlipCoin(node)}. 
In our experience, VERMEI independently reassigns the two pruning parameters (\(p_\text{parent}^d\) and \(p_{\text{leaf}}^d\)) before each pruning operation by sampling new values from a uniform distribution over [0.0, 1.0] as described in \cite{le2014compiler}. This controlled randomness helps increase the diversity of generated variants and improves the likelihood of exposing synthesis inconsistencies.

\subsection{Bug Identification}  
\label{Bug Identification}
 VERMEI automatically detects FPGA logic synthesis tool crash or unresponsiveness
(which we categorize as Hang/Crash Error) and only reports if it is
reproducible using the same Verilog variant. Besides, if the FPGA logic synthesis tool does not crash or time out, VERMEI identifes potential bugs through differential testing. We compare the synthesized output (\(O\)) of the seed Verilog test program against the outputs \((O_1, O_2, \dots, O_n)\) of the equivalent Verilog variants under the same testbench  conditions. If any synthesized variant produces a differing output from the original seed Verilog test program in terms of logic or timing behavior, it indicates a potential bug in the synthesis process.

\vspace{1.5em}
\section{Empirical Evaluation}
In this section, we evaluate the effectiveness of VERMEI. In particular, we seek to investigate the following research
questions (RQs):

\begin{itemize}
\setlength{\itemindent}{1em} 
\item[\textbf{RQ1:}] How is the defect-finding capability of VERMEI in FPGA logic synthesis tool  bugs?
\item[\textbf{RQ2:}] Can VERMEI detect more bugs in FPGA logic synthesis tools compared with state-of-the-art methods?
\item[\textbf{RQ3:}] Can the VERMEI enhance coverage improvement?
\item[\textbf{RQ4:}] How do the two mutation operations influence the performance of VERMEI on detecting FPGA logic synthesis tool bugs?
\item[\textbf{RQ5:}] How effective is the Bayesian sampling strategy of VERMEI?

\end{itemize}
In our experiments,  RQ1 and RQ2 are used to evaluate the bug finding capability of VERMEI compared to the state-of-the-art methods. RQ3 aims to evaluate the coverage improvement of VERMEI on logic synhesis tools. RQ4 is employed to assess the influence of two mutation operations. RQ5 aims to evaluate the sampling strategy of VERMEI. 

\subsection{Implementation}
We implemented VERMEI with approximately 2,000 lines of Python code and shell, the source code and experimental data are available at Figshare\footnote{https://figshare.com/s/17b24362c3258f05e942}. 
VERMEI uses an existing open-source fuzzer to generate seed Verilog programs, which are then mutated to detect logic synthesis bugs, following methodologies from prior compiler testing studies~\cite{tang2021detecting,guo2022detecting,jiang2021ctos,tu2022detecting,tu2023llm4cbi}. Specifically, we use seeds generated by Verismith or EvoHDL~\cite{Verismith}, ensuring that none of them trigger synthesis bugs initially. During preprocessing, we use design coverage information to identify and mark zombie logic.
We use VCS, a commercial simulator, to collect high-fidelity line coverage data for each test program, enabling the identification of zombie logic through unexecuted statements in the coverage output. These metrics are used only during preprocessing to validate inactive code regions with high precision~\cite{vcsfile}. For convergence analysis, we use Covered~\cite{Covered}, an open-source Verilog coverage analyzer, to evaluate structural coverage across large batches of automatically generated test programs. Covered provides line, branch, and condition coverage with minimal runtime overhead, making it suitable for large-scale comparative analysis.

For the implementation of bug identification, the differential testing is primarily performed (\sectionautorefname ~\ref{Bug Identification}). When a synthesis tool bug is detected, we employ a test program reduction process similar to the methodology described in related works such as Verismith \cite{Verismith}. This reduction process simplifies the Verilog test program that triggers the bug before it is reported on the developer's bug report website. This ensures that developers can quickly comprehend and resolve the identified bugs.

\subsection{Testing Setup}
 \textbf{Hardware.} Our evaluation is performed on two PCs with Intel(R) Core(TM) i7-8700 CPU @ 3.60GHz running Ubuntu 20.04 operating systems.  

 \textbf{FPGA logic synthesis tools.} In the study, we use three popular FPGA synthesis tools as subjects, namely Yosys, Xilinx’s Vivado, and Quartus Prime, following the existing synthesis tool testing studies \cite{Verismith}. Although previous studies also tested XST, we decided not to include it in our study because XST is part of the older Xilinx ISE tool suite, which has been discontinued and replaced by the more advanced Vivado tool.
 
 In RQ1 and RQ3, we rigorously tested the latest releases of three FPGA logic synthesis tools. Throughout the experiment, developers were actively updating their tools, which resulted in significant changes to the version. For instance, Yosys was updated from R0.39 to R0.40, with multiple iterations and improvements incorporated. This continuous evolution ensured that our testing was conducted on the most current and relevant versions of the tools, which provided up-to-date insights into their performance and capabilities. We chose these recent versions because developers primarily fix bugs in the latest releases rather than older versions. All identified bugs were reported to the respective bug tracking websites. 
 
 For RQ2 and RQ4, we utilize two versions of each 
 FPGA logic synthesis tool: Yosys (Yosys R0.15 and Yosys R0.40), Vivado (Vivado R2020.1 and Vivado R2023.2), and Quartus Prime (Quartus R20.1 and Quartus R23.1). These evaluated FPGA logical synthesis tools include both older release versions and recent release versions. The reason is that the older releases usually contain more bugs and give more statistically significant results, while the recent releases could check whether our method still works well. The choice of the older version (Yosys R0.15, Vivado Rv.2020.1, and Quartus R20.1) were chosen based on their similar release dates, which allowed for fair comparisons of different tools within similar technical time frames.

\textbf{Baseline approaches for RQ2 and RQ3}.
To illustrate the bug-finding capability of VERMEI, we compare VERMEI with the state-of-the-art method Verismith and EvoHDL. We selected EvoHDL as a baseline because it is currently the most representative structured random HDL test generation tool for testing FPGA logic synthesis tools. Additionally, we selected Verismith since, to the best of our knowledge, it is the first and currently the only work specifically designed to generate Verilog programs for the validation of FPGA logic synthesis tools.

\textbf{VERMEI varaint for RQ4.} To investigate whether the mutation operations contribute to VERMEI, we compare VERMEI with its two variants, i.e., VERMEI\textsubscript{\textit{prune}} and  VERMEI\textsubscript{\textit{insert}}. VERMEI\textsubscript{\textit{prune}} employs pruning operators to generate Verilog test program variants by pruning logic fragments from seed Verilog test programs. VERMEI\textsubscript{\textit{insert}}, on the other hand, only utilizes insertion operators to generate Verilog test program variants by adding new logic fragments within the regions of zombie logic.

\textbf{VERMEI sampling strategy for RQ5.} To evaluate the effectiveness of the Bayesian sampling strategy used in VERMEI for generating new logical fragments. We conducted an experiment comparing the mutation performance of VERMEI's optimized sampling strategy with that of a random selection strategy. The random selection strategy randomly selects logical statements or logic blocks from historical Verilog design files for mutation.

\textbf{Seed Verilog test programs.} We used Verismith to generate an initial set of  Verilog seed programs for VERMEI. These seed programs were manually or automatically verified to ensure that they do not trigger any synthesis bugs under our test setup. All selected seed programs produce consistent logic and timing behavior across multiple synthesis runs and tools, and none of them cause crashes or timeouts. This clean seed set ensures that any differential behavior detected by VERMEI arises from the mutations rather than pre-existing issues in the seeds. For each verified seed, we  generated 5 semantically equivalent variants,which is the same as ~\cite{guo2022detecting}. After each fuzzing round, newly generated test cases are added to the seed pool for further mutation. This iterative strategy allows the fuzzer to explore deeper semantic neighborhoods of promising seeds, increasing the chances of uncovering subtle synthesis inconsistencies. 

\textbf{Timeout setup.} During fuzzing, we monitor the execution time of each synthesis task to detect unresponsive behavior. For every generated Verilog variant, the synthesis tool is invoked in a subprocess with a fixed timeout threshold of 60 seconds, similar to the timeout strategy adopted in prior compiler fuzzing studies~\cite{chen2023compiler, cummins2018compiler}.
 If the synthesis process does not complete within this time window, it is forcibly terminated and the case is flagged as a potential Hang/Crash Error. We then manually analyze the tool’s logs and behaviors to distinguish between an actual crash (e.g., segmentation fault, internal error) and a hang (e.g., infinite loop or unresponsive state).
 This timeout value was determined empirically based on the observed runtime distribution of thousands of synthesis jobs, ensuring a balance between responsiveness and fairness.

\subsection{Answer to RQ1}
\begin{table*}[!t]
\centering
\caption{ VERMEI discoverred bugs}
\label{VERMEI discoverred bugs}
\footnotesize
\resizebox{0.9\textwidth}{!}{
\begin{tabular}{lllcc}
\toprule
 \# & TSC & Summary  & Status & Type \\
\midrule
1 & Yosys-\#4290  & Hang on a loop when write\_cxxrtl   & New    & H    \\
 2 & Yosys-\#3278   & Synthesis issue with shift (likely alumacc) & New  & M    \\
3 & Yosys-\#4336  & synth\_* passes should call check -mapped & New    & M    \\
 4& Yosys-\#3697 & Undef propagation in power operator& ?     & M    \\
5 & Yosys-\#4307  & Assertion Failure in genrtlil.cc with Signedness Issue Description  & New   & C    \\
6 & Yosys-\#3691  & Yosys fails in synthesizing D flip-flop   & Known   & M    \\
 7& Yosys-\#4335  & Assertion Failure in node-\&gt;bits == v at frontends/ast/ast.cc:855 & Known  & C    \\
8 & Vivado-LET7iSAH & Crash with HARTNDb::optimizeAndMap()  & New    & C    \\
9& Vivado-PKB7bSAH  & Bit-selection synthesis mismatch        & New    & M    \\
10 & Vivado-Qq9UsSAJ  & The signal bit width does not match       & Known   & M    \\
11& Vivado-Qq7fuSAB & Incorrect Synthesis for Nested Bitwise and Comparison Operations                     & New      & M    \\
12 & Vivado-Rqvc2SAB &Synthesis Mismatch with Conditional Selections and Shift Operations                               &  New      &  M    \\
13 &  Vivado-RhDHlSAN &The result of synthesis is unexpected and inconsistent with the original design                               &  New      &  M    \\
14 & Vivado-Qqw4ISAR &   Multimodule synthesis mismatch
  &   ?     &  M  \\
15& Quartus Prim-Question & Signed extension in the binary xor operation synthesis mismatch   & Known   & M  \\
\midrule
\end{tabular}}
  \begin{threeparttable}
 \begin{tablenotes}
 \footnotesize
\item[1]  The TSC is Technical Support Case ID from developers,There are three types of status feedback from developers on bug report (i.e., New = newly confirmed bug, Know = known bug, ? = under investigation). There
are three types of bugs (i.e., Type) in our reported bugs: hang bugs (H) crash bugs (C) and synthesis mismatches bugs (M)
\end{tablenotes}
 \end{threeparttable}
\end{table*}

\textbf{Methodology.} Detecting real bugs in mature FPGA logic synthesis tools is challenging, especially for commercial, well-established tools. To evaluate the bug detection capability of VERMEI in practice, we conducted a five-month experiment from December 2023 to May 2024. we rigorously tested the latest releases of three FPGA logic synthesis tools (i.e., Yosys R0.39 and Yosys R0.40 , Vivado Rv.2023.2, and Quartus R23.1). Because the developers actively updated their tools, Yosys was updated from R0.39 to R0.40 throughout our experiment. We chose these recent versions because developers primarily fix bugs in the latest releases rather than older versions. We submitted all the detected bugs as
issues to the bug report websites of various tool developers\footnote{https://figshare.com/s/704020b14d7a81df6912}.

\textbf{Result.} 
\tableautorefname ~\ref{VERMEI discoverred bugs} summarizes all the bugs detected in the latest development versions of the FPGA logic synthesis tools reported so far. There are three types of status of the feedback from developers on bug reports (i.e., New = newly confirmed bug, Know = known bug, and ? = under investigation ). The types of bugs we reported include hang bugs (H), crash bugs (C), and synthesis mismatches (M). Over a five-month period, we reported a total of 15 unique issues, which include 9 new bugs, 4 known bugs, and 2 bugs currently under investigation by the developers. Note that, as many  bugs have been fixed in the latest development versions of Yosys, Vivado, and Quartus, it has become increasingly challenging to detect new bugs in these updated versions of FPGA logic synthesis tools. Despite this, it is still worthwhile to detect new defects in the development versions to ensure the quality of FPGA logic synthesis tools.

\lstdefinestyle{customverilog}{
  language=Verilog,
  basicstyle=\ttfamily\scriptsize,
  keywordstyle=\color{darkred},
  commentstyle=\color{gray},
  numbers=left,
  numberstyle=\tiny\color{gray},
  stepnumber=1,
  numbersep=5pt,
  tabsize=2,
  showspaces=false,
  showstringspaces=false,
  breaklines=true,
  breakatwhitespace=true,
  captionpos=b,
  moredelim=**[is][\color{blue}]{@}{@},
  mathescape=false, 
  literate={\$}{{\$}}1 
}
\begin{figure}[t]
    \centering
    \begin{lstlisting}[style=customverilog, ]
if (wire3)
...
else
begin
    reg4 <= wire2;
    @reg5 <= (~|($unsigned($signed({wire3, reg4})) ?
    $unsigned((!{wire0, wire0})) : (reg4[(1'h1):(1'h1)] ?
    $unsigned($signed(reg4)) : ($signed(wire0) & (!(8'hbb))))));
    @ //generates statements
    reg6 <= $signed(reg4);
end
endmodule
    \end{lstlisting}
    \captionsetup{font=footnotesize} 
    \caption{TSC Yosys-\#4290. Hang with feedback arcs (\url{https://github.com/YosysHQ/yosys/issues/4290)}}
    \label{fig:Yosys4290}
\end{figure}

\begin{figure}[tbp]
    \centering
    \begin{lstlisting}[style=customverilog, ]
always @ (posedge clk or negedge rst_n)
begin
q <=!rst_n ? '0 : d;
end
endmodule\end{lstlisting}
 \captionsetup{font=footnotesize} 
    \caption{TSC Yosys-\#3691. Crash with flip-flop (\url{https://github.com/YosysHQ/yosys/issues/3691)} }
    \label{fig:Yosys3691.}
\end{figure}

\begin{figure}[!]
    \centering
    \begin{lstlisting}[style=customverilog, ]
module top (y, wire0, wire1, wire2, wire3);
 assign y = {wire4};
 assign wire4 = (+$signed((((~^wire3) >>> (wire0 >>> (8'hae))) ?
           ({wire3} << (-wire2)) : ((wire1 ?
             wire1 : wire3) * $unsigned(wire1)))));
endmodule \end{lstlisting}
    \captionsetup{font=footnotesize} 
    \caption{TSC Vivado-Rqvc2SAB. Synthesis Mismatch Due to Conditional Selections and Shift Operations (\url{https://support.xilinx.com/s/question/0D54U00008Rqvc2SAB)} }
    \label{fig:Vivado Rqvc2SAB}
\end{figure}

From \tableautorefname ~\ref{VERMEI discoverred bugs}, we find that developers have confirmed 1 hang bug, 3 crash bugs, and 11 incorrect synthesis bugs. \textbf{TSC Yosys-\#4290}, depicted in \figureautorefname ~\ref{fig:Yosys4290}, is a hang bug in Yosys R.39 identified by VERMEI. VERMEI creates assignment statements that form feedback loops through \texttt{adc.fir.ram.n980\_q} and \texttt{adc.fir.ram.n982\_q}, which triggers the bug. These feedback arcs prevent the  FPGA logic synthesis tool from finding a stable solution to the logic states, which causes Yosys to enter an infinite loop during synthesis. Crash bugs result in parsing errors or internal assertion failures in the FPGA logic synthesis tool.
For instance, \textbf{TSC Yosys-\#3691}, shown in \figureautorefname ~\ref{fig:Yosys3691.}, involves Yosys R0.40 failing to synthesize flip-flops with asynchronous reset and ternary operators during the \textit{PROC\_DFF} pass.
Specifically, Yosys tries to create a register for signal \texttt{\textbackslash ff2.\textbackslash q}, but the combination of asynchronous reset and ternary operators causes the synthesis to fail. Synthesis mismatch bugs in FPGA logic synthesis tools are more challenging to detect compared to hang and crash bugs. This difficulty arises because synthesis mismatches often manifest indirectly as logic failures within HDLs like VHDL or Verilog, rather than as direct and immediate errors. For instance, \textbf{TSC vivado-\#Rqvc2SAB}, illustrated in \figureautorefname ~\ref{fig:Vivado Rqvc2SAB}, is a synthesis mismatch caused by conditional selections and shift operations in Vivado R2023.2. VERMEI identified this bug, and a change request has been submitted to address it in a future release of Vivado.

However, existing methods (e.g., Verismith) have significant limitations in detecting these bugs. Under the same testing time, Verismith was only able to detect 3 known bugs in the latest version of Vivado, 1 known bug in the latest version of Yosys, and no bugs in Quartus. This limited performance can be attributed to Verismith’s restricted generation strategies. Verismith primarily constructs logical fragments based on Verilog syntax rules, which may limit its ability to cover the full spectrum of potential bugs since it focuses mainly on syntactical constructs. In contrast, VERMEI generates new program variants by altering control and data flows to test various optimization methods of FPGA logic synthesis tools. This method focuses on exploring a broader range of potential issues beyond basic syntax correctness. Furthermore, VERMEI employs a Bayesian sampling strategy, generating complex block sequences from historical data and existing patterns.

\textbf{Conclusion.} VERMEI is highly effective in detecting bugs in FPGA logic synthesis tools in practice. Over a period of five months, it reported a total of 15 valid bugs, of which 8
have been confirmed as new bugs.

\subsection{Answer to RQ2}
\textbf{Methodology.} To evaluate the effectiveness of VERMEI, we assessed its bug-finding capability against the state-of-the-art method Verismith, since the primary objective of these methods is to maximize the number of bugs found within a given time period. In our experiments, we allocated a single one week testing period for each version of FPGA logic synthesis tool. During this period, each method was used to test Yosys (versions Yosys R0.15 and Yosys R0.40), Vivado (versions Vivado R2020.1 and Vivado R2023.2), and Quartus Prime (versions Quartus R20.1 and Quartus R23.1).

To more clearly highlight the contribution of our mutation-based VERMEI, we used Verismith to generate only 1000 clean programs as the initial seed test cases for VERMEI during the one week fuzzing for each synthesis tool version. Under this setting, Verismith continues to generate a large number of test programs throughout the one week, significantly more than the 1000 initial seed test cases used by VERMEI. For EvoHDL, we selected an average CPS module count of 300 for test case generation, and we only generated Verilog-type test cases in our evaluation. This value was chosen within the configurable range of 100 to 600 modules supported by EvoHDL, aiming to balance structural complexity and generation efficiency through combinatorial construction. Over a one-week period, we generated test cases using this configuration for evaluation purposes. In this case, VERMEI does not incorporate too much from Verismith.  Hence, such experiments are useful to compare mutation-based VERMEI with syntax-based Verismith. For example, in one week fuzzing of Yosys R0.40, the Verismith tests generated 100,764 test cases during the process, which is significantly more than the 1,000 initial seed test cases used by VERMEI. VERMEI, on the other hand, we generated 120,080 test programs over the course of one week through mutation.
  Further, we analyze the bug-finding capability of VERMEI compared with the baselines from two aspects, i.e., the number of all detected bugs and the number of unique bugs. For duplicate bugs that occur across different versions of the same tool, we do not merge them; instead, we count each occurrence as a separate bug for measurement purposes. For example, if a bug in version 0.40 of Yosys also appears in version 0.15, we will record them separately, with each bug in the different versions counted individually.

\begin{figure*}[!t]
	\centering
	\includegraphics[width=0.9\linewidth]{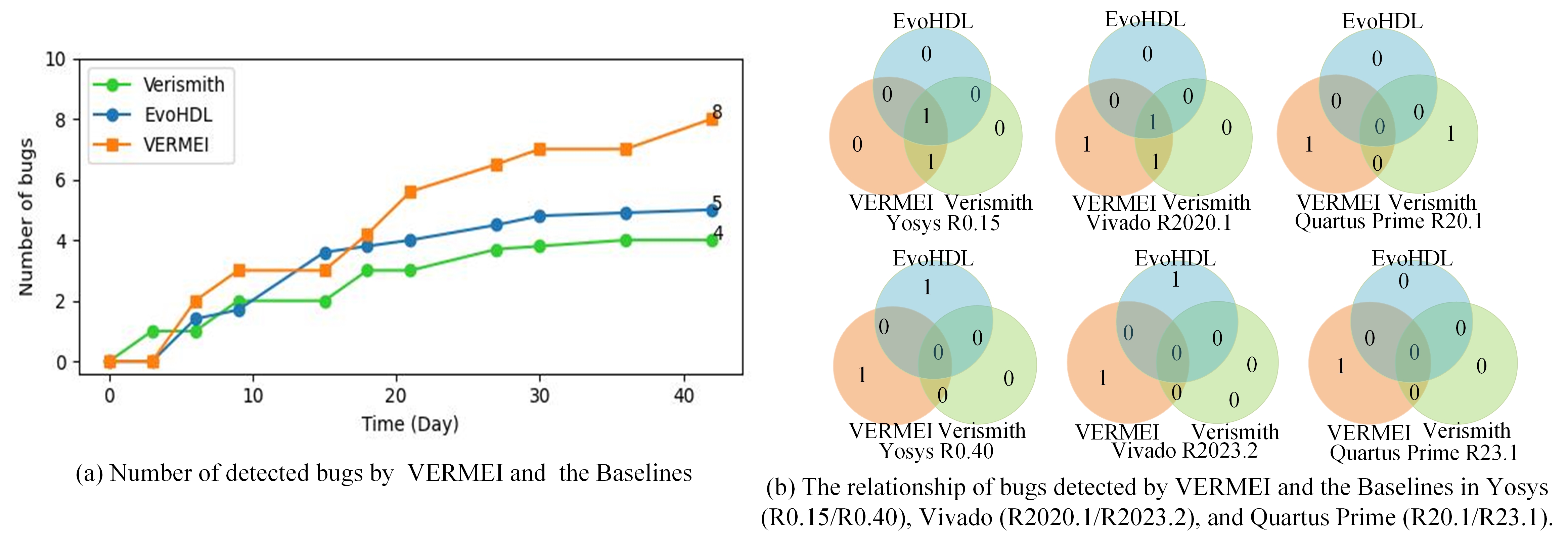}
	\caption{Comparison of Bugs Detected by  VERMEI the Baselines }
\label{The relationship of Bugs}
\Description{}
 \label{fig:The relationship }
\end{figure*}

\textbf{Result.}\figurename ~\ref{The relationship of Bugs}(a) shows the number of bugs detected by each technique during  total six weeks fuzzing period. Overall, VERMEI detected 8 bugs, while Verismith and EvoHDL detected 4 and 5 bugs, respectively, during the same period. This demonstrates that VERMEI achieved a 100\% improvement in bug detection effectiveness compared to Verismith and a 60\% improvement compared to EvoHDL. Verismith and EvoHDL's bug detection rate quickly reaches saturation, while VERMEI continues to detect bugs throughout the fuzzing process. This disparity stems from the limited structural and semantic diversity in the test case generation methods employed by Verismith and EvoHDL. Specifically, Verismith uses fixed grammar-based rules for test generation, which leads to highly similar test cases that fail to explore diverse synthesis paths. EvoHDL generates test cases by combining its inherent CPS modules, the limited number and types of these modules restrict the structural and semantic diversity of the generated test cases.

Furthermore, we analyzed the performance of VERMEI, Verismith, and EvoHDL in detecting bugs across different types of FPGA logic synthesis tools, and the results are summarized in \tableautorefname ~\ref{Bugs detected}. \tableautorefname ~\ref{Bugs detected} classifies the detected bugs into newly discovered bugs (New) and previously known bugs (Known), where “Known” refers to duplicates of bugs already documented in the existing bug repository. As shown in \tableautorefname ~\ref{Bugs detected}, VERMEI detected a total of 8 bugs across six versions of three different FPGA logic synthesis tools, including 5 new bugs and 3 known bugs.  In contrast, Verismith detected a total of 4 bugs, with 1 new bug and 3 known bugs, while EvoHDL detected 5 bugs in total, including 2 new bugs and 3 known bugs. Moreover, to show the relationship between the bugs detected by  VERMEI and the baselines, we depict this relationship in \figurename ~\ref{The relationship of Bugs}(b). From \figurename ~\ref{The relationship of Bugs}(b), we can see that VERMEI always detects the largest number of unique bugs. The reason is that VERMEI can generate  semantically rich and complex variants with distinct control flow and data flow characteristics, which lead to different structural and signal path properties. When an FPGA logic synthesis tool synthesizes a Verilog test program, the synthesizer employs static analysis to identify optimization techniques appropriate for the current control flow and data flow patterns. VERMEI can exercise FPGA logic synthesis tool more thoroughly by forcing it to use various optimization methods on variants. 

\begin{table}[!t]
\captionsetup{font=footnotesize} 
\caption{Bugs detected by VERMEI and the baselines in a week fuzzing campaign per version of Yosys (R0.15/R0.40), Vivado (R2020.1/R2023.2), and Quartus Prime (R20.1/R23.1).}
\label{Bugs detected}
\resizebox{0.5\textwidth}{!}{
\begin{tabular}{lcccccccccccccc}
\toprule
&&\multicolumn{2}{c}{Verismith} & \multicolumn{2}{c}{EvoHDL}&\multicolumn{2}{c}{VERMEI} \\
\cmidrule(r){3-4} \cmidrule(r){5-6} \cmidrule(r){7-8} 
& & New & Known & New & Known& New & Known \\
\midrule
Yosys R0.15 & &0 &1& 1 &1  & 1 &1 \\
Yosys R0.40 &&0 & 0& 0 &1 &1 &0 \\
Vivado R2020.1 &&1 &1& 1 &0&1 &1 \\
Vivado R2023.2 &&0 &0& 0 &1&1 &0\\
Quartus Prime R20.1 &&0 &1& 0 &0 &0 &1 \\
Quartus Prime R23.1 && 0&0& 0 &0 &1 & 0\\
\midrule
\textbf{Total} &&1&3&2&3 &5 &3 \\
\bottomrule
\end{tabular}}

\end{table}

\textbf{Conclusion.} The experimental results demonstrate that VERMEI significantly outperforms Verismith and EvoHDL for testing FPGA logic synthesis tools. VERMEI detected a total of 8 bugs across six versions of three different FPGA logic synthesis tools, with 5 of these are new bugs. In contrast, Verismith and EvoHDL reached saturation in bug detection relatively quickly, while VERMEI can detect bugs continuously within the fuzzing time. These  results verify the bug-finding capability of VERMEI.

\subsection{Answer to RQ3}
\label{sec:convergence analysis}
\textbf{Methodology.} Test coverage  provides insights into the thoroughness of the testing process \cite{coverage,sun2016finding}. Test coverage metrics, such as line, branch, and condition coverage, indicate how much of the Verilog code has been executed during testing. Higher coverage generally means that more code paths have been tested, which helps in identifying potential bugs that might otherwise remain undetected. To evaluate the coverage improvement of VERMEI on logic synhesis tools in comparison with the state-of-the-art methods (Verismith and EvoHDL).
We have chosen Yosys as our FPGA logic synthesis tool because it is the only available open-source option suitable for this purpose. 
Coverage evaluation is conducted using using 
Covered - Verilog Code Coverage Analyzer \cite{Covered}. Covered primarily supports line coverage, condition coverage and branch coverage. The line coverage measures whether each line of Verilog code is executed by the test programs \cite{ammann_offutt_2008}. For example, it checks if a register assignment statement has been executed during the testing process. Condition coverage measures whether all possible outcomes of a conditional expression are covered. For example, it checks if the boolean condition in an if statement evaluates to both true and false. Branch coverage measures whether each branch of branch statements (such as if-else and case statements) has been executed.
 Therefore, in this experiment, we aim to compare the improvements in line, condition, and branch coverages achieved by VERMEI with those obtained using Verismith and EvoHDL. Our baseline is the coverage obtained by executing 10,000 test programs. Specifically, Verismith and EvoHDL randomly generates 10,000  Verilog test programs, whereas VERMEI generates 10,000 mutants by applying mutations to seed programs (each seed generating only one mutant).  
 To ensure a fair comparison of coverage improvements between VERMEI and the baselines, the 10,000 test programs of VERMEI do not include the initial seed programs.

\begin{figure*}[!t]
	\centering
	\includegraphics[width=0.9\linewidth]{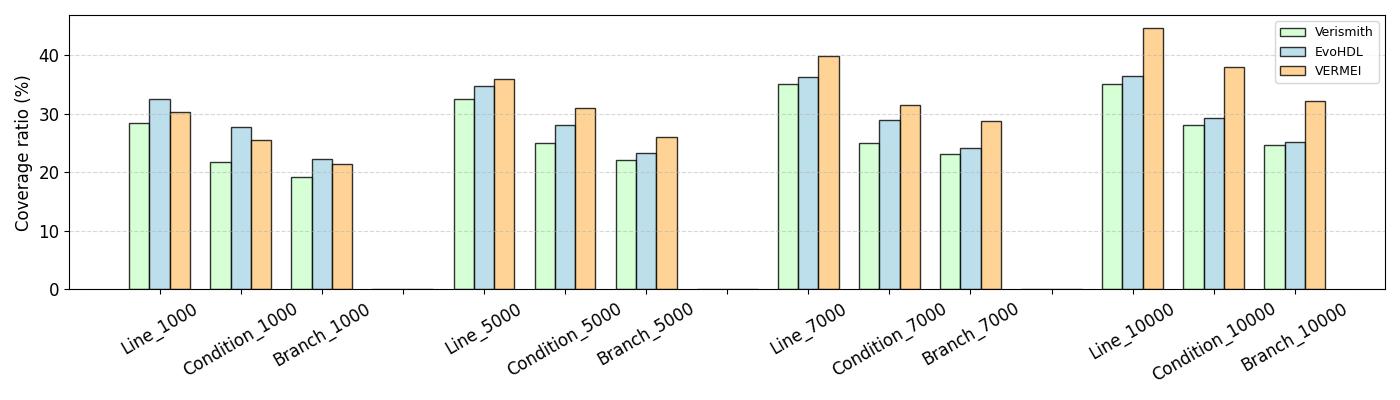}
	\caption{ The coverage of VERMEI and the baselines on line, condition and branch. }
 \label{coverage}
\end{figure*}

\textbf{Result.} \figurename ~\ref{coverage} shows the coverage  of VERMEI and Verismith on line, condition, and branch. VERMEI achieved 44.64\% line coverage, 38.04\% condition coverage, and 32.20\% branch coverage. In comparison, Verismith achieved 35.04\% line coverage, 28.14\% condition coverage, and 24.68\% branch coverage, while EvoHDL achieved 36.50\% line coverage, 29.30\% condition coverage, and 25.10\% branch coverage. Overall, VERMEI improves coverage over Verismith by 9.60\% (line), 9.90\% (condition), and 7.52\% (branch), and over EvoHDL by 8.14\% (line), 8.74\% (condition), and 7.10\% (branch), respectively. Although EvoHDL employs structure-aware test generation and achieves slightly higher coverage than Verismith, it still lags behind VERMEI across all coverage metrics under the same testing budget of 10,000 test programs. This is primarily because EvoHDL relies on a fixed set of CPS modules (with at most around 600 modules) to generate Verilog programs, which fundamentally limits the diversity of its generated programs. The limited number of available modules and their composition patterns constrain the structural and semantic diversity in the generated tests.

While VERMEI and Verismith both contribute to increasing line, condition, and branch coverages, VERMEI demonstrates clear superiority over Verismith and EvoHDL. Notably, as the number of test programs increases beyond 5,000, VERMEI consistently achieves higher coverage across all metrics compared to both baselines, and its coverage continues to increase with the addition of more test programs. In contrast, the coverage improvements of Verismith and EvoHDL tend to plateau as the number of test programs grows, indicating a saturation point in their ability to explore new execution paths. This phenomenon is expected.
 Random generation (Verismith) and structure-aware generation (EvoHDL) can quickly cover common and shallow paths when the number of test cases is small. However, after these easy-to-trigger paths are covered, new test cases often trigger the same paths repeatedly, causing the coverage to stop increasing. In contrast, VERMEI uses mutation-based test generation. It mutates seed programs to change their semantics and control flows. This helps VERMEI trigger new branches and conditions that other tests miss, even as the test suite grows. 

\textbf{Conclusion.}VERMEI outperforms both Verismith and EvoHDL in improving line, condition, and branch coverage. Specifically, it achieves increases of 9.60\%, 9.90\%, and 7.52\% over Verismith, and 8.14\%, 8.74\%, and 7.10\% over EvoHDL, respectively. These improvements are expected, given VERMEI's capability to generate a more diverse set of Verilog test programs.
These improvements are as expected due to its ability to generate a more diverse set of Verilog test programs.

\subsection{Answer to RQ4}
\textbf{Methodology.}
To investigate whether the mutation operations contribute to VERMEI, we compare VERMEI with its two variants, i.e., VERMEI\textsubscript{\textit{prune}} and  VERMEI\textsubscript{\textit{insert}}.  For VERMEI and its two variants,  we run each version of the 
 FPGA logic synthesis tool 10 times, with a timeout of 72 hours, as in the previous study \cite{tang2021detecting}. This results in a total of 90 experiments (three comparative methods, 10 runs for each FPGA logic synthesis tool). Notably, different FPGA logic synthesis tools have distinctive bug exposure patterns in test environments, and even under the same test conditions, different tools or versions may show unique bug exposure circumstance due to different internal optimizations. In this RQ, there are no bugs were found in Quartus R 22.1.1 and Quartus R 23.1 by VERMEI and the corresponding comparison methods during the given test period. So,we do not report test results for Quartus R 22.1.1 and Quartus R 23.1. To effectively utilize CPU computation resources, we ran many experiments simultaneously on each PC, allowing us to complete the experiments in approximately 45 days. During each experiment, VERMEI\textsubscript{\textit{prune}} and VERMEI\textsubscript{\textit{insert}} are configured to generate the same number of variants for each seed Verilog test program (i.e., 10 variants  in our experiments).


\begin{table*}[t]
\centering
\caption{Number of detected bugs for VERMEI and its variants}
\label{VERMEI and its variants}
\resizebox{0.98\textwidth}{!}{
\begin{tabular}{ccccccccccccccc}
\toprule
\multirow{2}{*}{\textbf{Subject}}  & \multicolumn{2}{c}{\textbf{VERMEI}} & \multicolumn{5}{c}{VERMEI\textsubscript{prune}} & \multicolumn{5}{c}{VERMEI\textsubscript{insert}} \\
\cmidrule(r){2-3} \cmidrule(r){4-8} \cmidrule(r){9-13}
   & \#Avg.  &\#TP. & \#Avg. &\#TP.& imp. & P-value & Effect size & \#Avg. &\#TP. & imp. & P-value & Effect size \\
\midrule
Yosys R0.15 & 4.8 & 143.39 & 2.7 &153.21  & 77.78\% & $<$0.001 & 1.000 & 3.9&132.68 & 23.08\% & $<$0.001 & 0.955 \\

Yosys R0.40 & 3.2 & 150.54 & 2.3&164.21& 39.13\% & $<$0.001& 0.980  & 2.9 &139.36& 10.34\% & 0.002 & 0.860 \\


Vivado R2020.1 & 3.6 & 141.15 & 2.6 &141.78 & 38.46\% & 0.001 & 0.940 & 3.2 &138.34 & 12.50\% & 0.001 & 0.940 \\

Vivado R2020.1  & 2.4 & 124.87 & 1.7 & 128.36 & 41.18\%  & $<$0.001 & 0.905 &2.1&120.23 & 14.28\% & 0.001 & 0.835 \\

\midrule
Total & 14.0 & 559.95 & 9.31 & 587.56 & 50.37\% & - & - & 12.1 & 530.61 & 15.70\% & - & - \\

\bottomrule
\end{tabular}}
\end{table*}

\begin{figure*}[!t]
	\centering
	\includegraphics[width=1\linewidth]{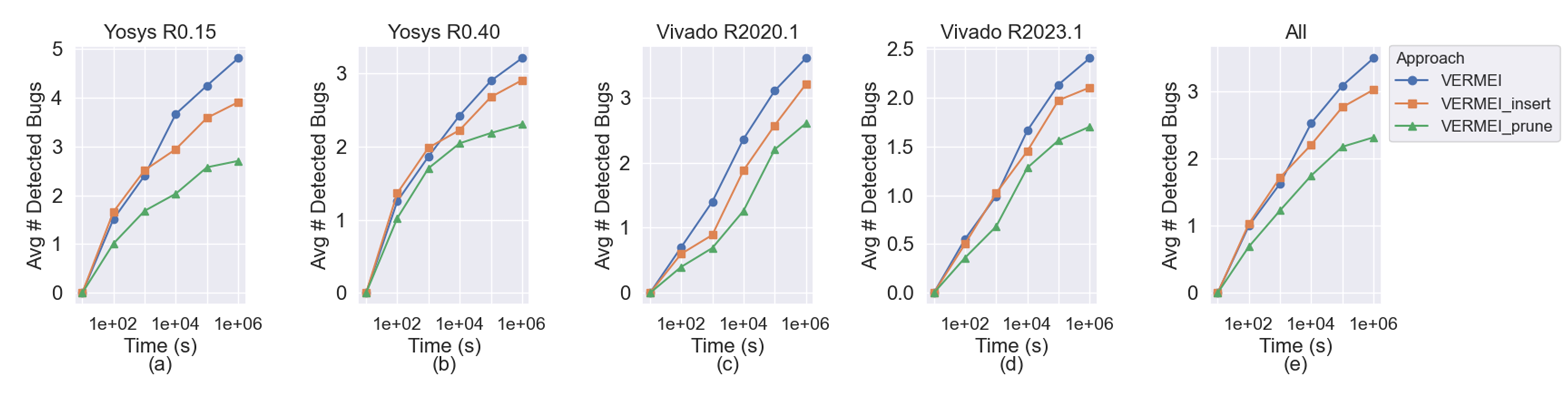}
	\caption{Time spent on detecting bugs for VERMEI and its variants}
 \Description{}
 \label{fig:Time}
\end{figure*}

\textbf{Result.} In this subsection, we first describe the quantitative results of the reported bugs and then show the time spent detecting the bugs in each test scenario of VERMEI and its variants.

\textit{1)}\textbf{\textit{ quantitative results}.} This subsection
describes some statistical properties of the discovered bugs, including  the average number of defected bugs (i.e., \#Avg. bugs), the average number of valid
test programs (i.e., \#TP.) and improvement (i.e., \( \text{imp} = \frac{VERMEI - baseline}{ baseline} \times 100\% \)) of VERMEI over its two variants on the average number of bugs and the statistics results of Mann-Whitney U-test \cite{arcuri2011practical} for each run. \tableautorefname ~\ref{VERMEI and its variants} presents the average number of bugs identified by VERMEI and its variants over 10 runs. From these results, it is evident that VERMEI outperforms its variants in bug detection. Specifically, across all experiments, VERMEI detects an average total of 14.0 bugs, while VERMEI\textsubscript{\textit{prune}} and VERMEI\textsubscript{\textit{insert}}  identify 9.31 and 12.1 bugs, respectively, representing improvements of 50.37\% and 15.70\%.
In particular, the improvement columns \(imp.\) of VERMEI\textsubscript{\textit{prune}} and VERMEI\textsubscript{\textit{insert}}  in \tableautorefname ~\ref{VERMEI and its variants}  reveal that VERMEI detects more bugs than VERMEI\textsubscript{\textit{prune}} and VERMEI\textsubscript{\textit{insert}}  by margins of 16.95\% \textasciitilde 32.83\% and 12.63\% \textasciitilde 21.68\%, respectively. This suggests that the integration of both mutations in VERMEI generates more bug-sensitive structures compared to utilizing a single operation. Notably, VERMEI\textsubscript{\textit{insert} }demonstrates better effectiveness in bug detection than VERMEI\textsubscript{\textit{prune}}. This indicates that inserting mutation  constructs more bug-sensitive structures than merely pruning mutation from the test programs.

Additionally, the P-value (p$<$0.05) in \tableautorefname ~\ref{VERMEI and its variants} demonstrates that VERMEI significantly outperforms its variants. Furthermore, we calculated the effect size of the differences between VERMEI and the variants using the Mann-Whitney U-test. We also calculate the effect size of the differences between VERMEI and the baselines using the Vargha and Delaneys A12 statistics \cite{A12,tang2021detecting,zhou2022locseq}. 
According to this test, if VERMEI and the baselines are equivalent, then A12 = 0.5; if the effect of VERMEI is small compared to the baselines, then A12 $<$ 0.5; otherwise, A12$ >$ 0.5. As shown in Table 2, all the effect sizes are greater than 0.7 which indicates that VERMEI has a higher probability to obtain better results than its variants.

\textit{2)} \textbf{\textit{time spent on detecting bugs}.} \figurename ~\ref{fig:Time}(a)-(e) shows the time required for bug detection by VERMEI and its variants, with \figurename ~\ref{fig:Time}(e) providing the overall results. The data in clearly demonstrate that VERMEI consistently outperforms its variants in detecting the average number of bugs in both Yosys and Vivado. For example, the overall results in \figurename ~\ref{fig:Time}(e) reveal that VERMEI detects an average of 2.53 bugs within $10^4$ seconds.
In contrast,  VERMEI\textsubscript{\textit{prune}} detects 1.75 bugs, and VERMEI\textsubscript{\textit{insert}} detects 2.21 bugs in the same period. 
This corresponds to improvements of 44.57\% and 14.48\%, respectively. Particularly, in \figurename ~\ref{fig:Time}(c), VERMEI detects an average of 1.5 bugs over $10^3$ seconds, while VERMEI\textsubscript{\textit{prune}} and VERMEI\textsubscript{\textit{insert}} detect only 0.8 and 0.7 bugs, respectively, during the same testing period. Additionally, VERMEI spends less time detecting the same number of bugs compared to its variants. For example, in \figurename ~\ref{fig:Time}(e), VERMEI detects an average of 2.5 bugs within $10^2$ seconds, whereas VERMEI\textsubscript{\textit{prune}} takes nearly 2 magnitudes of the time on detecting  same number of bugs.

\textbf{Conclusion.} VERMEI exhibits a higher probability of achieving superior results compared to its variants.
 On average, VERMEI detects a total of 14.0 bugs, whereas VERMEI\textsubscript{\textit{prune}} and VERMEI\textsubscript{\textit{insert}}  identify 9.31 and 12.1 bugs, respectively. Additionally, VERMEI spends less time detecting the same number of bugs compared to its variants, which demonstrates greater efficiency in detecting bugs in FPGA logic synthesis tools.

\subsection{Answer to RQ5}
\begin{figure}[!t]
	\centering
	\includegraphics[width=1\linewidth]{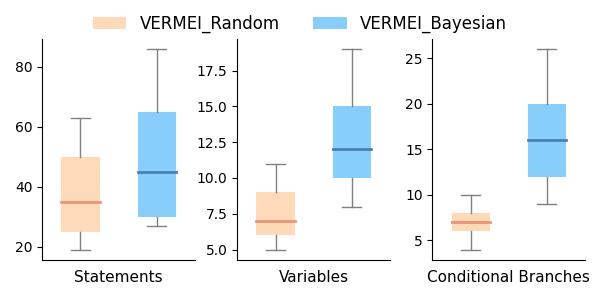}
	\caption{The distribution of generated Verilog variants by  VERMEI\_Bayesian and VERMEI\_Random.}
 \Description{8.}
 \label{fig:sample}
\end{figure}

\textbf{Methodology.} To evaluate the effectiveness of the Bayesian sampling strategy used in VERMEI for generating variant designs, we conducted an experiment comparing the mutation performance of VERMEI's optimized sampling strategy with a random selection strategy. The random selection strategy is to randomly select logic statements or logic blocks from the historical Verilog design file datas for mutation operations. In this Case, we tested the variants generated by VERMEI using Bayesian sampling (VERMEI\_Bayesian) and random sampling (VERMEI\_Random) by analyzing the characteristics of the resulting variant files. Specifically, we counted the total number of Verilog variants generated over one week by both approaches when testing Yosys R0.40. The box-plots in \figurename ~\ref{fig:sample} show the collected three metric values, including the number of newly added statements, newly introduced variables, and new conditional branches in each variant Verilog. For conditional branches, we focused on explicit conditional structures, including \textit{if-else}, \textit{case}, \textit{for}, and ternary operators \textit{(? :)}.

\textbf{Result.} Number of newly added statements, variables, and conditional branches. The number of newly added statements, variables, and conditional branches can intuitively reflect the complexity of the design file. After the design files have been synthesized into RTL netlists, the increase in variables means that it is more likely that complex gates will be combined. And the increase in conditional branches may introduce more control paths and states.
As shown in \figurename ~\ref{fig:sample}, we can see that the minimum and maximum number of newly added statements, variables, and conditional branches in the variant files generated by VERMEI\_Bayesian are [27, 86], [8, 19], and [9, 26], respectively. The results show that the  new logical fragments of the variant test cases generated by VERMEI\_Bayesian increase the overall complexity of the Verilog test case to a certain extent. 

\textbf{Conclusion.} The effectiveness of VERMEI's Bayesian sampling strategy outperforms the random sampling method.  Bayesian sampling method can generate diverse and complex new logical fragments in Verilog variants with higher semantic logic complexity, which can partially explain why VERMEI can effectively detect bugs in FPGA logic tools.


\section{Threats Validity }
\subsection{Internal Threats.} The threats to internal validity mainly lie in the implementations of VERMEI. As described in \sectionautorefname ~ \ref{sec:apporach}, the effectiveness of VERMEI hinges on accurately identifying and managing zombie logic.Zombie logic refers to code segments that remain inactive under certain input conditions and clock cycles. If zombie logic is incorrectly identified, it may lead to unintended modifications of active logic or missed mutation opportunities, reducing the effectiveness and correctness of the generated test cases. To address this threat, we use the widely adopted VCS~\cite{vcs} tool for coverage-based zombie logic annotation, ensuring precise detection of inactive code segments. We also design targeted testbenches to enhance coverage and improve the detection of zombie logic. Additionally, we perform extensive simulations and validate against benchmark designs to confirm that zombie logic is correctly identified and handled within VERMEI, reducing the risk of semantic inconsistencies during mutation. 

\subsection{Extenal Threats.} A key threat to the external validity of VERMEI is the limited range of subjects used in our experiments. We tested only six versions of the Yosys, Vivado, and Quartus Prime logic synthesis tools, including three older releases and three recent ones. This selection may not adequately represent other logic synthesis tools,which makes it unclear if our approach can be generalized to different HDL synthesizers and simulators. To address this issue, future research will involve applying VERMEI to a wider variety of synthesizers and simulators for fuzz testing, thereby enhancing the robustness and applicability of our methodology across diverse HDL synthesis environments.

\section{Relation Work}
\subsection{FPGA logic synthesis tool testing.}
In the realm of FPGA synthesis compilers testing, Verismith~\cite{Verismith} and EvoHDL~\cite{ledohdl} represent two prominent approaches for testing FPGA synthesis compilers.  Verismith~\cite{Verismith}, It operates as a Verilog program generator, generates pseudo-random, valid, and deterministic Verilog test programs using an AST-based approach. However, Verismith’s generator often produces simple test programs, mainly covering basic combinational logic, making it hard to reflect the complex structures in industrial designs.  EvoHDL~\cite{ledohdl} leverages CPS models to generate diverse HDL code in Verilog, VHDL, and SystemVerilog, providing a broader range of inputs for testing synthesis tools. However, the HDL generated by EvoHDL is somewhat limited, as it primarily focuses on supporting multiple hardware description languages rather than providing meaningful semantic diversity within each language. To address these limitations and enhance the thoroughness of FPGA logic synthesis compilers testing, we introduced VERMEI, aiming to ensure the accuracy and correctness of FPGA synthesis processes.

 Another tool is VlogHammer~\cite{VlogHammer},VlogHammer is a Verilog code generator based on define and sample a subset of Verilog syntax. VlogHammer generates non-deterministic Verilog code, cannot generate programs containing multiple modules, and does not support behavioral-level Verilog (e.g., always blocks). 

 Additionally, generators like VERGEN~\cite{vergen} produce behavioral-level Verilog by randomly combining high-level logic blocks such as state machines, multiplexers, and shift registers. However, they rely on predefined structures, which limits the diversity of generated programs and reduces their ability to cover a wide range of Verilog designs.

Recently, Large Language Models (LLMs) have been widely used in test case generation. The same methods are employed in FPGA logic synthesis compilers testing. The most utilized method, VeriGen~\cite{vergen}, a prominent approach, fine-tunes the CodeGen-16B model and improves the syntactic correctness of generated Verilog code by 41\% over pretrained models like GPT-3.5-turbo. Other approaches, such as BetterV~\cite{BetterV}, mainly focus on optimizing Verilog code and reducing runtime.  However, existing LLM-based methods show promise in Verilog code generation, it is not specifically focused on testing.

Researchers have also worked to improve FPGA compiler reliability. He Jiang~\cite{DeLoSo}introduced DeLoSo to detect bugs during complex netlist optimizations. Yann Herklotz et al.~\cite{9786208} proposed Vericert, a formally verified high-level synthesis (HLS) tool. Zun Wang et al.~\cite{wangzun} used fuzz testing with various C programs to thoroughly evaluate HLS tools.

Despite these contributions, there remains a gap in the quality  and complex structures in industrial designs of the generated files. Consequently, we have embarked on a more comprehensive testing of FPGA logic synthesis compilers to affirm their stability and reliability.


\subsection{ Differential testing and Equivalence checking}
 Differential testing has become  the standard
method for checking the correctness. This technique involves compiling the same input program with multiple compilers and comparing the outputs \cite{mckeeman1998differential,tu2022detecting,tang2021detecting,guo2022detecting}. If the outputs differ, it is presumed that a bug exists in at least one of the compilers. Tools like DIFUZZRTL \cite{DIFUZZRTL} leverage this method to automatically discover unknown bugs in CPU RTLs.

In hardware design, FPGA logic synthesis tools convert high-level descriptions into lower-level representations like Verilog, and a distinct method is required to ensure correctness. Unlike relying on differential testing alone, a method that may not be practical given the diversity of 
 FPGA logic synthesis tools, formal equivalence checking is often used. \cite{goldberg2001using,karfa2008equivalence,ramos2011practical}.
The equivalence checking mathematically verifying that the synthesized output (e.g., netlist) is functionally equivalent to the original design specification, which enables the detection of any differences introduced during synthesis \cite{molitor2007equivalence,barrow1984verify}. Commercial tools like Conformal \cite{Cadence} and Synopsys Formality \cite{synopsys} are widely used for logical equivalence checking. These tools are designed to handle the complexities of modern hardware designs, comparing netlists to the original RTL descriptions. By providing a rigorous framework for equivalence checking, these tools help ensure that the synthesized hardware faithfully implements the intended design without introducing errors, thereby enhancing the reliability of hardware synthesis processes.

However, equivalence checking is not always suitable for testing FPGA logic synthesis tools. While modern equivalence checking methods are typically based on SAT/SMT solvers, they may still encounter limitations when dealing with extensive structural transformations or unmodeled behaviors introduced during synthesis. Specifically, the structural transformations and optimizations introduced during synthesis often go far beyond simple combinational modifications. In the absence of additional information, equivalence checking tools may produce false positives, or fail to complete the verification process due to state space explosion when dealing with large-scale sequential designs.

\section{Conclustion and Future Work}
In this paper, we propose VERMEI, a new differential testing with equivalent variants method for detecting deeper bugs in FPGA logic synthesis tools. VERMEI consists of three modules: preprocessing, equivalent mutation, and bug identification. First, the preprocessing module identifies zombie logic in seed test programs through simulation and coverage analysis. Then, the equivalent mutation module generates equivalent variants  by iteratively pruning or inserting new logic fragments in the zombie logic areas.This module uses Bayesian sampling to extract logic fragments (e.g., complex control flow structures, nested conditional statements, and intricate computation structures) from historical Verilog  designs for insertion. This makes the generated equivalent variants more complex, with intricate control flows and structures, thus addressing the challenge of test programs that have insufficient semantic logical complexity.  Finally,
the bug identification module employs differential testing based on the equivalent variants to verify all Verilog test program variants. 
Our evaluation demonstrates that VERMEI significantly outperforms existing the state-of-the-art methods. Within five months, we reported 15 bugs to vendors, 9 of which were confirmed as new bugs.

The main limitation of VERMEI lies in its restriction of mutation operations to the zombie logic regions, which significantly narrows the available mutation space and makes it challenging to generate sufficiently rich and diverse mutation effects. 

In future work, we plan to enhance VERMEI by expanding the variant space and increasing the diversity of mutations. Additionally, we aim to devise more efficient fuzzing techniques for FPGA logic synthesis tools to further improve their testing and reliability.



\bibliographystyle{IEEEtran}
\bibliography{refs}



\end{document}